\documentclass[useAMS,usenatbib]{mn2e}

\usepackage{bm}
\usepackage{amsmath}
\usepackage{amssymb}
\usepackage{color}
\usepackage{pst-grad}
\usepackage{psfrag}
\usepackage{graphicx}

\newcommand{\mnras}{MNRAS}
\newcommand{\apj}{ApJ}
\newcommand{\apjs}{ApJS}
\newcommand{\apjl}{ApJ}

\newcommand{\aap}{A\&A}

\newcommand{\eqb}{\begin{eqnarray}}
\newcommand{\eqe}{\end{eqnarray}}
\newcommand{\eqbn}{\begin{eqnarray*}}
\newcommand{\eqen}{\end{eqnarray*}}
\newcommand{\diff}{{\rm d}}
\newcommand{\pdiff}[2]{\frac{\partial #1}{\partial #2}}

\newcommand{\im}{\rm i}

\title[Universal behaviour of cosmic-ray mediated precursors]
{Universal behaviour of shock precursors in the presence of efficient cosmic-ray acceleration}

\author[Reville \& Bell]{B. Reville\thanks{E-mail:
b.reville1@physics.ox.ac.uk}, A.~R. Bell\\
Clarendon Laboratory, University of Oxford, Parks Road, Oxford OX1 3PU, United Kingdom}

\begin{document}

\date{Accepted \dots Received \dots}

\pagerange{\pageref{firstpage}--\pageref{lastpage}} \pubyear{2525}

\maketitle

\label{firstpage}

\begin{abstract}
The self-consistent interaction between energetic particles and self-generated hydromagnetic waves
in a cosmic-ray pressure dominated plasma is considered.
Using a three-dimensional hybrid MHD-kinetic code, which utilises
a spherical harmonic expansion of the Vlasov-Fokker-Planck equation, high resolution
simulations of the magnetic field growth including feedback on the cosmic rays
are carried out. It is found that for shocks with high cosmic-ray acceleration 
efficiency, the magnetic fields become highly disorganised, resulting in
near isotropic diffusion, independent of the initial orientation of the ambient magnetic field.
The possibility of sub-Bohm diffusion is demonstrated for parallel shocks, 
while the diffusion coefficient approaches the Bohm limit from below 
for oblique shocks.
This universal behaviour suggests that Bohm diffusion in the 
root mean squared 
field inferred from observation may provide a realistic estimate for the
maximum energy acceleration timescale in young supernova remnants. Although disordered,
the magnetic field is not self-similar suggesting non-uniform energy dependent behaviour
of the energetic particle transport in the precursor. Possible indirect radiative 
signatures of cosmic-ray driven magnetic field amplification are discussed. 
\end{abstract}


\section{Introduction}

There is a growing wealth of observational evidence that
non-linear amplification of magnetic fields occurs at the outer shocks of 
supernova remnants, both upstream and down
\citep{achterberg94,vinklaming03,bambaetal05,rakowskietal11}. This discovery
represents a significant development in the theories of 
particle acceleration and the resulting non-thermal emission from supernova 
remnants. However, despite the fact that almost
every attempt to model the non-thermal spectra from young supernova remnants 
in recent years has incorporated or inferred some level of field amplification,
a detailed understanding of the underlying mechanisms that produce such
strong fields is still lacking. This can probably be attributed to the complexity
of the problem, which, being highly non-linear, requires numerical modelling 
of processes occurring on a variety of length-scales.

The amplification of magnetic fields to levels far in excess of typical
interstellar medium values is advantageous for a number of reasons. 
First, it seems essential to account for the x-ray synchrotron emission, its short 
time scale variability, and its localisation to narrow filaments close to the 
shock. Second, it is central to explaining the origin of cosmic rays with 
energies in excess of $10^{15}$eV. While the former can in principle be 
accounted for via fluid instabilities occurring at the shock itself and
the region immediately downstream \cite[e.g.][]{guoetal}, the latter 
requires strong fields upstream of the shock that can only be produced 
from instabilities driven by energetic particles streaming ahead of the shock
\cite[e.g.][]{achterberg83,bell04}.
While the non-resonant hybrid (NRH) instabilities described in \cite{bell04,bell05} 
are believed to have the fastest growth rates, whether
the resulting short-wavelength structures can significantly influence 
the highest energy particles has yet to be
verified \citep{bell04,revilleetal08}. 

In this paper, we report on numerical simulations of the interaction 
between streaming cosmic rays and the thermal plasma upstream of the shock, 
including a self-consistent treatment of the cosmic rays.
The thermal plasma is treated using an MHD fluid approach while the cosmic ray 
evolution is calculated using a spherical harmonic expansion of the 
Vlasov-Fokker-Planck equation. 

The outline of the paper is as follows. In the next section we introduce the 
relevant equations and the spherical harmonic expansion that is central to 
the approach used in this paper. From the resulting equations, the 
standard expressions for particle drifts and fluxes in the diffusion approximation
are derived. These
form the basis for many of the approximations required to self-consistently
model the essential 
processes occurring in the precursor of an efficiently accelerating shock. 
Section 3 describes 
the numerical scheme and the simulation results are presented in section 4.
We conclude with a summary of the main conclusions and additional discussion.

\section{Particle transport and fluxes}
\label{sims_sect}

The study of particle acceleration and magnetic field amplification in
supernova precursors is essentially one of magnetohydrodynamics. Due to the
large length-scales associated with the energetic cosmic rays, their interactions
with the background field are mediated by the magnetic fluctuations supported by
the background plasma. However, it is the currents provided by the cosmic-rays 
that often play the dominant role in determining the evolution of the magnetic 
field in these precursors.
We seek an effective numerical method to model the evolution of the cosmic rays 
in self-generated hydromagnetic waves and structures. For non-relativistic 
shocks $u_{\rm sh}\ll c$, it is generally accepted that efficient scattering 
maintains a quasi-isotropic phase space distribution, at least to order 
$u_{\rm sh}/c$. In a previous paper \citep{revillebell12}, a particle in cell
type numerical scheme was adopted to treat the cosmic ray evolution 
\cite[see also][]{zacharythesis,lucekbell00}. Maintaining an isotropic distribution 
in multi-dimensions turned out to be computationally expensive, with the 
number of particles per cell required to achieve the 
desired level of isotropy becoming prohibitively large for simulations in three 
spatial dimensions.

Recently \cite{Belletal11} performed test particle 
simulations of cosmic-ray acceleration at oblique shocks in one dimension.
The numerical scheme involved a spherical harmonic expansion of the distribution 
in momentum space based on the KALOS code previously developed for 
simulations of laser plasma interactions \citep{kalos}. 
In this paper, we describe the full three dimensional expansion, which has been 
coupled to an MHD code. We adopt the mixed-frame coordinate system commonly 
used in astrophysics, where momentum variables are measured in the local 
fluid frame 
\eqb
\label{vfp_local}
&&\pdiff{f}{t}
+(\bm{u}+\bm{v})\cdot \bm{\nabla}f
-p\bm{E}\cdot\pdiff{f}{\bm{p}}\nonumber\\
&&-\left[(\bm{p}\cdot\bm{\nabla})\bm{u}\right]\cdot\pdiff{f}{\bm{p}}
+e\bm{v}\cdot\left(\bm{B}\times\pdiff{f}{\bm{p}}\right)=
\mathcal{C}(f)\enspace,
\eqe
accurate to second order in $u/c$ \citep{saasfee}. In this non-inertial 
frame an effective electric field term is introduced \cite[cf.][]{skilling75}:
$$\bm{E}=\frac{1}{c}\left[\pdiff{\bm{u}}{t}+(\bm{u}\cdot\bm{\nabla})\bm{u}\right].$$ 
We have also kept the third order term $(\bm{u}\cdot\bm{\nabla})\bm{u}$ simply 
because the resulting expression for the electric field can be inferred from the 
momentum equation of magnetohydrodynamics,
which turns out to be convenient when included in the code.
$\mathcal{C}(f)$ represents the usual Fokker-Planck small angle scattering
term \cite[e.g.][]{chandrasekhar43}. The advantage of using the
mixed coordinates system is that the resulting collisions, as given by 
$\mathcal{C}(f)$, are elastic and isotropic in the local frame, 
in the sense that they are proportional to the angular component of the 
Laplacian in spherical coordinates. Since the spherical harmonics are themselves
solutions to Laplace's equation, the resulting collision term takes a very simple form.

Defining  $f_\ell^{-m} = (f_\ell^{m})^*$ the expansion takes the following form:
\footnote{For a more rigorous discussion of this expansion see \cite{tzoufras}
and references therein.}
\begin{equation}
f(p,\theta,\varphi) = \sum_{\ell=0}^\infty\sum_{m=-\ell}^\ell f_\ell^m(p) 
P_\ell^{\vert m\vert}(\cos\theta) e^{im\varphi}
\end{equation}
which, on substitution into equation (\ref{vfp_local}), can be used to derive 
a set of coupled differential equations. These equations are quite
lengthy, and for clarity they are written in single column form in
the appendix. 

To motivate the various approximations that are made later in the simulations,
we demonstrate how the well known transport equation is reproduced from the 
full expansion. For this purpose, it is necessary to retain only terms $\ell<2$. 
In the full numerical simulations, many more terms are used.

Defining $f=f_0+\bm{f_1}\cdot\bm{p}/p$ where 
\eqbn
f_0=f_0^0, \enspace f_x=f_1^0,\enspace  f_y=-2\Re f_1^1, \enspace f_z = 2\Im f_1^1
\eqen
and neglecting the terms associated with $\bm{E}$ above (which are second order in 
$u/c$), a somewhat tedious calculation shows that:
\begin{flalign}
\label{eqn:f0}
&\pdiff{f_0}{t}+\bm{u}\cdot\bm{\nabla}f_0+\frac{v}{3}\bm{\nabla}\cdot\bm{f_1}
-\frac{1}{3}(\bm{\nabla}\cdot\bm{u})p\pdiff{f_0}{p}
=0 \enspace ,&\\
\label{eqn:f1}
&\pdiff{\bm{f_1}}{t}+(\bm{u}\cdot\bm{\nabla})\bm{f_1}+v\bm{\nabla}{f_0}
+\bm{\Omega}\times\bm{f_1} = \frac{\bm{\nabla}u_i}{5p^3}
\frac{\partial(p^4f_i)}{\partial p}& \\
&\enspace
+\frac{1}{5}\pdiff{\bm{u}}{x_i}
p^2\frac{\partial}{\partial p}\left(\frac{f_i}{p}\right)
+\frac{1}{5}(\bm{\nabla}\cdot\bm{u})
p^2\frac{\partial}{\partial p}\left(\frac{\bm{f_1}}{p}\right)-\nu \bm{f_1}\enspace ,&
\nonumber
\end{flalign}
where $\bm{\Omega}=e\bm{B}/\gamma mc$ is the directional rotational frequency,
$\nu$ is the collision frequency and summation over repeated index 
$i=x,y,z$ is implied. In what follows we assume 
all particles are ultra-relativistic, and set everywhere $v=c$. 
Making the usual assumption
that the first order terms react on a shorter timescale than the zeroth 
order component, we take a steady state solution for $\bm{f_1}$. 
Retaining only terms to first order in $u/c$ leaves
\eqbn
c\bm{\nabla}{f_0}
+\bm{\Omega}\times\bm{f_1} = -\nu \bm{f_1}\enspace,
\eqen
which can be used to remove the $\bm{f_1}$ dependence in the equation (\ref{eqn:f0})
to give 
\begin{flalign}
\label{eq:full}
 &\pdiff{f_0}{t}+\bm{u}\cdot\bm{\nabla}f_0
=\frac{1}{3}(\bm{\nabla}\cdot\bm{u})p\pdiff{f_0}{p}&\\
& \enspace +\bm{\nabla}\cdot\left\lbrace
\frac{D_{\rm B}}{(1+h^2)}\left[
\bm{\nabla}f_0 + \bm{h}\left(\bm{h}\cdot\bm{\nabla}f_0\right) 
-\bm{h}\times\bm{\nabla}f_0 \right] \right\rbrace\enspace,& \nonumber
\end{flalign}
where $\bm{h}=\bm{\Omega}/\nu$ is the hall parameter, and $D_{\rm B} = c^2/3\nu$
the usual Bohm diffusion coefficient. 
It can clearly be seen that in the relevant limits, the transport equation reduces to 
the well-known forms for parallel shocks
\eqb
\label{eq:difflmt}
\pdiff{f_0}{t}+\bm{u}\cdot\bm{\nabla}f_0
=\frac{1}{3}(\bm{\nabla}\cdot\bm{u})p\pdiff{f_0}{p} +\bm{\nabla}\cdot
\left(D_{\rm B}\bm{\nabla}f_0\right) 
\eqe
and perpendicular shocks
\eqbn
\pdiff{f_0}{t}+\bm{u}\cdot\bm{\nabla}f_0
=\frac{1}{3}(\bm{\nabla}\cdot\bm{u})p\pdiff{f_0}{p} +\bm{\nabla}\cdot
\left[D_{\bot}(\bm{\nabla}-\bm{h}\times\bm{\nabla})f_0\right]\enspace, 
\eqen
where $D_{\bot}=D_{\rm B}(1+h^2)^{-1}$ \citep{bell08}.

We seek steady-state solutions, which will later be used to 
initialise the simulations. In the shock frame, the upstream plasma
moves in the positive $x$ direction with uniform speed $u_{\rm sh}$, 
and only gradients in this direction are considered. Without loss of
generality, a steady magnetic field is taken to lie in the $x-z$ plane 
(i.e. $\bm{h} = h_x \bm{\hat{x}}+h_z \bm{\hat{z}}$).
The resulting upstream steady state solution is
\eqb
\label{CRgrad}
f_0(x,p) = f_0(0,p)\exp\left(\int \kappa \diff x\right),
\mbox{~with~} 
\kappa = \frac{u_{\rm sh}}{D_{\rm B}}\frac{1+h^2}{1+h_x^2}
\eqe
and the associated fluxes 
\eqb
\label{CRfluxes}
\bm{f_1}(x,p) = \frac{3u_{\rm sh}}{c}
\left[-\bm{\hat{x}} + \frac{h_z}{1+h_x^2} \bm{\hat{y}} -
\frac{h_x h_z}{1+h_x^2}\bm{\hat{z}}
\right] f_0(x,p)\enspace .
\eqe 

In the limit that the amplitude of the scattering waves (or $\nu$)
does not increase significantly over a distance $\sim \kappa^{-1}$, the 
results reduce to equations (5) and (6) of \cite{Belletal11}. 
However, small wavelength modes are in principle excited by the 
NRH instability of \cite{bell04}, with maximum
growth rate (for quasi-parallel shocks)
\eqbn
\gamma_{\rm nr} = \frac{1}{2}\frac{n_{\rm cr}}{n_{\rm th}}
\frac{u_{\rm sh}}{v_{\rm A}}\Omega 
\eqen
such that the number of exponential growth factors in amplitude of
magnetic fluctuations over a distance 
equal to the fluid crossing time of $\kappa^{-1}$ is 
\eqbn
\gamma_{\rm nr}/\kappa u_{\rm sh} = 
\frac{n_{\rm cr}}{n_{\rm th}}
\frac{c}{u_{\rm sh}} \frac{c}{v_{\rm A}} 
\frac{h}{6}\frac{1+h_x^2}{1+h^2}
\eqen
For efficient shock acceleration, this number can exceed unity. In fact, at the 
highest energies, it must exceed unity if magnetic fields are to be amplified 
upstream.

\section{Numerical scheme}

The numerical scheme is a natural extension of earlier hybrid
MHD-kinetic work based on the MHD code used in \cite{revilleetal08}, with 
the particle-in-cell approach of \cite{revillebell12} replaced by a 
Vlasov-Fokker-Planck routine. The full expansion of equation
(\ref{eqn:A1}) is solved numerically using a third order Runge-Kutta method. 
Making use of the orthogonality relations of the spherical harmonics, it follows 
that the cosmic-ray particle and current density, measured in the local fluid frame, 
are
\eqb
n=q\int\diff^3 p f = 4\pi\int\diff p\, p^2 f_0^0
\eqe
and 
\eqb
\bm{j}'_{\rm cr}=q\int\diff^3 p \, \bm{v} f = \frac{4\pi}{3} q\int\diff p\,p^2 v \left(
\begin{array}{c}
f_1^0 \\ -2{\Re(f_1^1)} \\ {2\Im(f_1^1)}
\end{array}\right)\enspace.
\eqe

The MHD equations, modified by the presence of cosmic rays have been described in 
detail elsewhere \citep{zacharythesis,bell04}, 
but we repeat them here for completeness.
The standard MHD momentum equation is supplemented with the additional effect of the 
return current induced by the cosmic-ray streaming, giving
\begin{eqnarray*}
\rho\frac{\diff\bm{u}}{\diff t} = - \bm{\nabla} p  
-\frac{1}{4\pi}\bm{B}\times(\bm{\nabla}\times \bm{B}) -\frac{1}{c}
\bm{j}'_{\rm cr} \times \bm{B} \enspace.
\end{eqnarray*}
Defining the MHD energy density $e=\rho u^2 /2 + P/(\gamma-1) +B^2/8\pi$, 
where $\gamma$ is the adiabatic index of the background gas, it follows that the
energy density also differs from its usual conservative form
\eqb
\pdiff{e}{t} + \nabla \cdot
\left[ \left(e+p+\frac{B^2}{8\pi}\right)\bm{u} - 
\frac{1}{4\pi}( \bm{u} \cdot\bm{B})\bm{B}\right]=-\bm{j}'_{\rm cr}\cdot\bm{E}
\eqe
i.e. if the background plasma can set up an electric field to oppose
the cosmic ray current, work is done by the cosmic rays on the MHD fluid.
Note that there is no need to transform back to the laboratory 
frame since the return current is always defined locally to oppose that 
of the cosmic rays. 

Since the primary focus of this paper is the growth of magnetic waves 
and its feedback on the particles driving the growth, we make the following 
approximations:

\begin{enumerate}
 \item To reduce memory requirements and allow for a larger spatial domain as 
well as larger number of harmonics, we dispense with the momentum grid. This is 
achieved applying the same technique used in \cite{Belletal11}, where a uniform 
power-law distribution is assumed for all harmonics 
\eqb
\pdiff{f_{\ell}^m}{p} = -s\frac{f_{\ell}^m}{p}  
\eqe
This is similar to making a mono-energetic approximation, but 
ensures the $\bm{E}\times\bm{B}$ drifts remain properly accounted for.
For the parameters we consider, this is the most important effect. To prevent 
divergence of the cosmic-ray pressure, we have taken the value $s=4.1$ in all 
simulations. This also ensures all $H_3^m$ terms (see appendix, equation 
\ref{eq:GandH}) do not vanish exactly.

\item a finite collision frequency $\nu (\ll\Omega)$ is chosen in order to guarantee 
the closure of the spherical harmonic expansion. As can be seen from 
the appendix (Eq. \ref{app:C}), in the absence of other effects 
the high harmonics are damped exponentially
\eqbn
f_{\ell}^m(t+\Delta t) =f_{\ell}^m(t) 
\exp\left(-\frac{\nu}{2}\ell(\ell+1)\Delta t\right)\enspace.
\eqen
It is therefore possible to truncate the harmonic expansion  
at $\ell = L_{\rm max}$, subject to the condition that $L_{\rm max}>\sqrt{\Omega/\nu}$.
It will be demonstrated that after the field has been allowed to grow 
sufficiently, the self-consistent scattering on magnetic field structures dominates 
the artificial collisions imposed by $\nu$.

\item Perhaps the most crucial approximation that we adopt, which is a 
theoretical tool quite unique to the spherical harmonic expansion technique, is
to add an additional term proportional to the large scale gradient of the isotropic
part of the spectrum $f_{\rm LS}(x)$. This results in an additional source term
in the $A_{x,1}^0$ expression for $f_1^0$ (Equation \ref{eq:advect}), 
i.e. we include 
$$\hat{A}_{x,1}^{0} = -c \pdiff{f_{\rm LS}}{x}\enspace.$$
This term acts like an external driving force, which allows us to model the larger 
scale dynamics of the precursor. In the absence of self-generated magnetic fields,
this term would allow the system to relax to the standard steady state solution,
where advection and diffusion balance. In order to use this approximation,    
the gradient must vary on scales larger than the length of the
simulation domain $L_{\rm sim}$, ie.
$$ f_{\rm LS}\left/\left|\pdiff{f_{\rm LS}}{x}\right|\right. \gg L_{\rm sim}$$
Similarly, the use periodic boundary conditions for all other quantities demands
that this condition be satisfied, since otherwise the large scale gradient is 
resolved, and different boundary conditions or full shock simulations are necessary.
The gradient that we use in the simulations is taken to be that associated with the steady state
solution described in the previous section, where the value of $D_{\rm B}$
is determined from the numerical value for $\nu$ used in the simulation.

\end{enumerate}

Previous PIC/Hybrid kinetic simulations of cosmic-ray streaming instabilities
including feedback on the driving particles, have 
suffered from the fact that the energy must be transferred to the background 
plasma at the expense of the cosmic-ray anisotropy
\citep{lucekbell00, stromanetal09, revillebell12}. Hence, only a finite 
amount of free energy, determined by the initial conditions, 
was available in the system to amplify the magnetic field.
\cite{bell04} circumvented this problem by noting that on scales much less than 
the gyroradius of the cosmic rays, the feedback on the particles was, at least in
the linear theory, negligible, and a constant uniform current could be used.
In this paper we bridge the gap between these two regimes, by using the 
local approximation above, which allows us to control the driving of the 
cosmic ray current.

\section{Simulation results}

Despite the fact that a spherical shock expanding into a large scale 
uniform magnetic field has considerably 
more of its shock surface at large obliquities, parallel or quasi-parallel shocks
have received by far the most attention. This may be largely due to the 
belief that injection into the acceleration process is suppressed at high Mach number
quasi-perpendicular shocks \cite[e.g.][]{baringetal}. 
However, observations of supernovae would suggest that
this is not the case. Shell-type supernovae are quite abundant in our galaxy,
and it is extremely unlikely that these rims represent exclusively quasi-parallel
shocks. Kinetic simulations of the shock microphysics are providing new
insight \cite[see e.g.][and references therein]{gargate12, amano}, 
but the solution to the injection problem in its entirety remains unsolved. 
In the following, we assume that the shock has already put a considerable 
amount its energy into the cosmic-ray population, but not yet significantly 
amplified the magnetic fluctuations.

The simulations are performed in three dimensions, using periodic boundary 
conditions, where the computational domain represents 
the evolution of a small region at rest in the precursor of a supernova remnant shock.
Using the local approximation described in the previous section, we investigate
the effect of the angle between the shock normal, as represented by the large 
scale gradient, and the mean magnetic field. A turbulent component, 
$\langle\delta B^2 \rangle/B_0^2 \approx 1\%$, is added to the background magnetic
field, to seed the growth of waves. This seeding is essential as VFP codes do not 
suffer from noise. All simulations begin with the plasma completely at rest, and
plasma beta of unity. While the simulations attempt to model as close as possible 
the parameters relevant for young supernovae, the separation of scales is always a 
challenge. This manifests itself through the different timescales for the MHD updates,
determined by the maximum fast-magnetosonic velocity, and the cosmic-ray 
velocity, which is essentially the speed of light. Thus the CFL condition on the 
VFP equation requires a certain amount of sub-cycling.
To allow the simulations to run 
in a reasonable time we are forced to use an initial Alfv\'en velocity
somewhat larger than expected in a real situation, 
typically $\lesssim 10^{-3}c$. Thus, most of the simulations consider shocks 
with Alfv\'en Mach numbers on the order of $M_{\rm A} \sim 50-100$. The code is  
terminated when the maximum fast magnetosonic speed approaches the speed of light,
which usually results due to rapid non-adiabatic heating in the non-linear regime.
This non-adiabatic heating may have observational signatures which will 
be discussed in section \ref{sect:reg}.

\subsection{Oblique shocks}

The velocity of the intersection point between the shock and a given magnetic
field line is $u_{\rm int} = u_{\rm sh}/\cos \theta_N$, where $\theta_N$ is 
the angle between the large scale field and the shock normal. If the fields 
are strongly amplified to values $\delta B/B_0 > 1$, this quantity is, 
at least locally, quite poorly defined. However, since we start with quite modest 
fluctuations, we refer to the obliquity with regards initial mean field. 
Already at $u_{\rm sh}=c/10$ an angle $\theta_N > 85^\circ$ is required for 
a superluminal intersection velocity, such that even at quite large obliquities,
particles should be able to escape upstream along fieldlines. 

\begin{figure}
 \includegraphics[width=.45\textwidth]{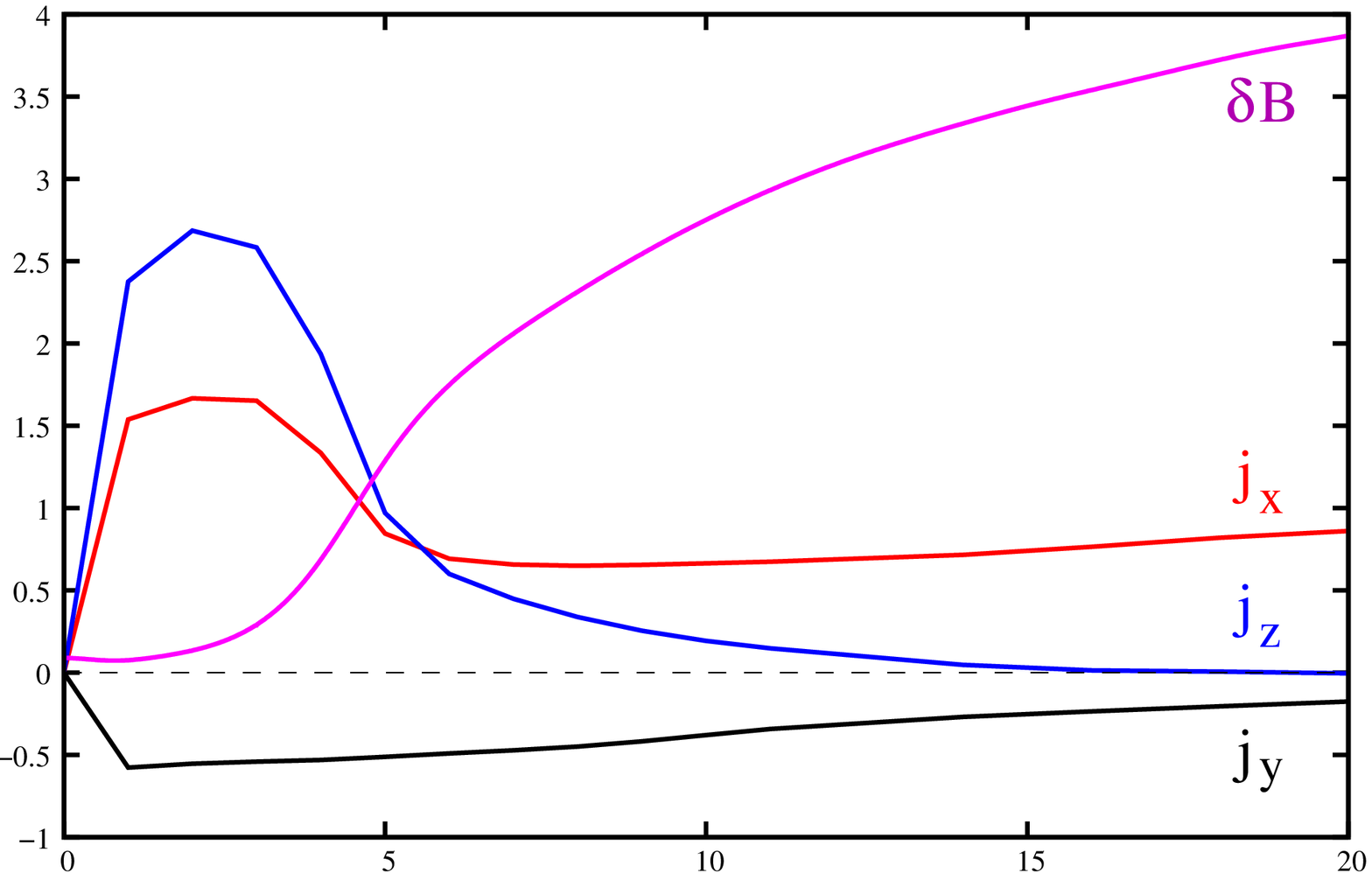}
 \includegraphics[width=.45\textwidth]{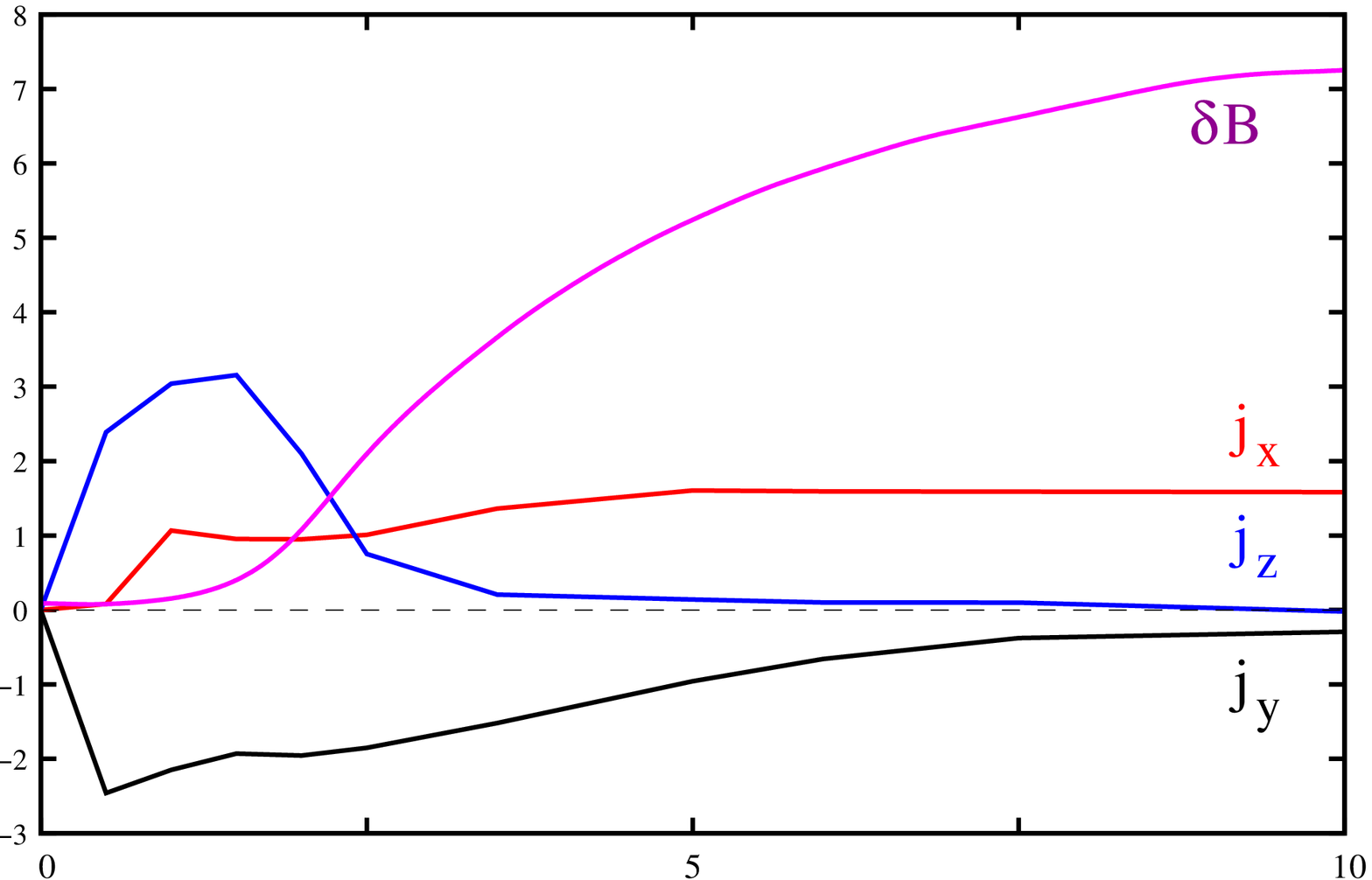}
 \includegraphics[width=.45\textwidth]{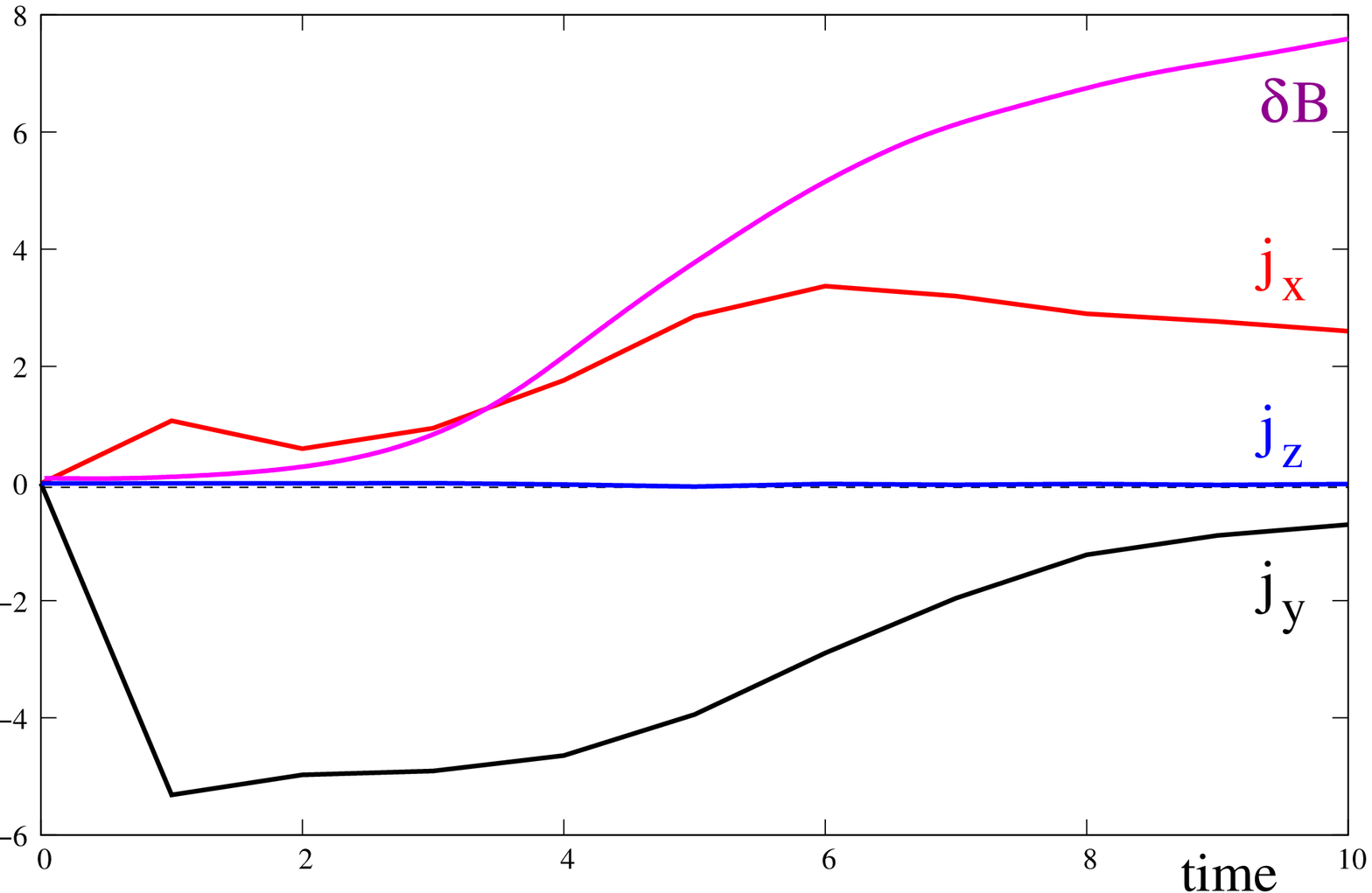}
\rput(-6.6,15.){$(a)~~\theta_N=60^\circ$}
\rput(-6.6,9.8){$(b)~~\theta_N=80^\circ$}
\rput(-6.6,4.6){$(c)~~\theta_N=90^\circ$}
 \caption{The lab frame current densities $\bm{j}=n_{cr}e\bm{u}+\bm{j}'$ 
and growth of magnetic field as 
a function of time. Initial conditions are chosen such that
$\nu = 0.1 \Omega$, $u_{\rm sh} = 0.03 c$.
Magnetic field is in units of $B_0$.
Time units are $20.8/\Omega_0$ i.e. if the CR are at 100 TeV, and
$B_0=3~\mu$G, time is approximately in years.
}
\label{fig:angles}
\end{figure}

Starting with the smaller obliquities, figure \ref{fig:angles} (a) shows the time 
evolution of the fluxes and turbulent field component for a shock at an 
initial angle $\theta_N=60^\circ$ to the shock normal. 
The cosmic-ray fluxes grow quickly from zero to match the values predicted from
Eq. (\ref{CRfluxes}). In fact the peak values correspond almost exactly to these
values, suggesting that the diffusion approximation is reasonable at this point, 
and that the scattering is not yet dominated by the self generated fields. 
This is to be expected since flux-freezing determines that the fields cannot 
grow before the plasma is set in motion, so that there is an initial inertial 
phase where no growth is observed \citep{bell04,revilleetal08}, and the cosmic-ray
scattering is likewise unaffected in the weak fluctuations. However, once
the field starts to grow, the magnitudes of all components of the cosmic-ray 
current begin to decrease. By the time the field becomes completely disorganised 
$\delta B \sim B_0$, the shock begins to behave gradually more like a 
parallel shock, for which the parallel component of the current dominates. This
parallel component continues to grow on account of the strong gradient which assumes
a fixed value for $\kappa$ (equation \ref{CRgrad}) relevant for a 
$60^\circ$ angle between the shock normal and the magnetic field. 

Increasing the magnetic field angle to $80^\circ$, it can be seen that
the time evolution of the various components of the current changes slightly,
Figure \ref{fig:angles} (b). The 
parallel component is found to grow slowly at first, but then overshoots
the value predicted by the diffusion approximation, Eq. (\ref{CRfluxes}).
The overshoot corresponds approximately to the fluctuation level where 
the self-consistent scattering dominates ($\delta B/B_0 > \nu/\Omega$), 
at which point the cosmic rays diffuse across the field lines more easily.
Previous VFP simulations performed by \cite{hornsby}, investigating
cross field transport in the presence of synthetic turbulence, found that
cross field transport approaches the Bohm limit at $\delta B \sim 2 B_0$.
Returning to the results shown in Figure \ref{fig:angles} (b), it is found that
the value for $j_y$ also overshoots, while $j_z$ falls short of its
predicted value. This is also consistent with a very rapid increase in the collision 
frequency. This may be due in part to the zero-th order $\bm{j}_{\rm cr}\times\bm{B_0}$
force accelerating the plasma almost instantly as the angle grows. 
The currents appear to saturate in a very similar fashion to the previous run at 
$60^\circ$ (compare Figures \ref{fig:angles} (a) and \ref{fig:angles} (b)). 

A similar behaviour occurs at the extreme limit of superluminal shock propagation,
$\theta_N=90^\circ$, where both the $x$ and $y$ components of the 
current overshoot, with $j_x$ gradually increasing into the non-linear regime before 
saturating, while $j_y$ decreases towards zero, Figure \ref{fig:angles} (c). 
Again, the driving term is fixed in 
time so the gradient is very steep. The regulation of this gradient is discussed
further in section \ref{sect:reg}

In all cases, the final stage of evolution appears to be similar, namely that 
cosmic rays are prevented from sliding along the field lines, as represented by 
$j_z$, while the drifts orthogonal to the plane of the mean field are gradually
damped. This suggests a universal behaviour for efficiently 
accelerating shocks, such that they tend asymptotically towards parallel-like 
shocks. While this phenomenon has been widely believed to occur, 
this is the first time that it has been demonstrated in a self-consistent manner. 
The global effect this has on the acceleration is uncertain at present.

A possible by-product of this behaviour may reveal itself on interaction 
with the shock surface itself. Previous numerical 
simulations by \cite{zirakashvili} demonstrated that for a parallel 
shock, the non-linear structures generated by the NRH instability
can, on crossing the shock, result in stretching of the magnetic component 
parallel to the shock normal, such that it dominates over the component in the shock
plane. We find that similar structures are found in simulations at all obliquities, 
suggesting that even initially perpendicular
shocks may result in radially pointing magnetic fields downstream of young 
supernova remnant shock, as has been observed in several remnants \citep{Milne}.

A caveat regarding the non-linear evolution concerns the finite lifetime 
of the simulation region, being in the upstream rest-frame, it must 
at some stage be advected through the shock.
This problem is most severe in the case of 
highly oblique shocks, where the precursor is
initially on the order of only a few gyroradii. However, as the magnetic field
becomes increasingly disordered, particles will penetrate deeper into the 
upstream plasma, and allow further time for magnetic field growth. Ultimately,
for shocks efficiently accelerating cosmic-rays, self-generated fields take over
and the shock will behave more like a parallel shock. However, sufficiently 
far upstream, the fields will remain regular, at least at the level of pre-existing
interstellar turbulence, and particle escape to infinity is impeded. 
Clearly full shock simulations 
are required to treat this problem self-consistently. This will be addressed in 
a follow-up paper.

It is tempting to connect these results to the previous work of 
\cite{Belletal11}, where it was demonstrated that the shape of the 
non-thermal power law could deviate from the standard test-particle 
value at oblique shocks, depending on the angle and collision frequency.
This deviation is essentially a direct consequence of the inability to match 
the perpendicular currents across the shock in the diffusion approximation,
which demands $f_0(x>0,p)={\rm const}$. We have 
just demonstrated that the time-asymptotic 
state of any efficiently accelerating shock front is one in which the perpendicular
currents vanish, which implies the existence of a possible radiative signature
that can be used to probe the level of upstream field amplification. To
illustrate this point, SN 1987a is taken as an example.

\begin{figure*}
 \includegraphics[width=\textwidth]{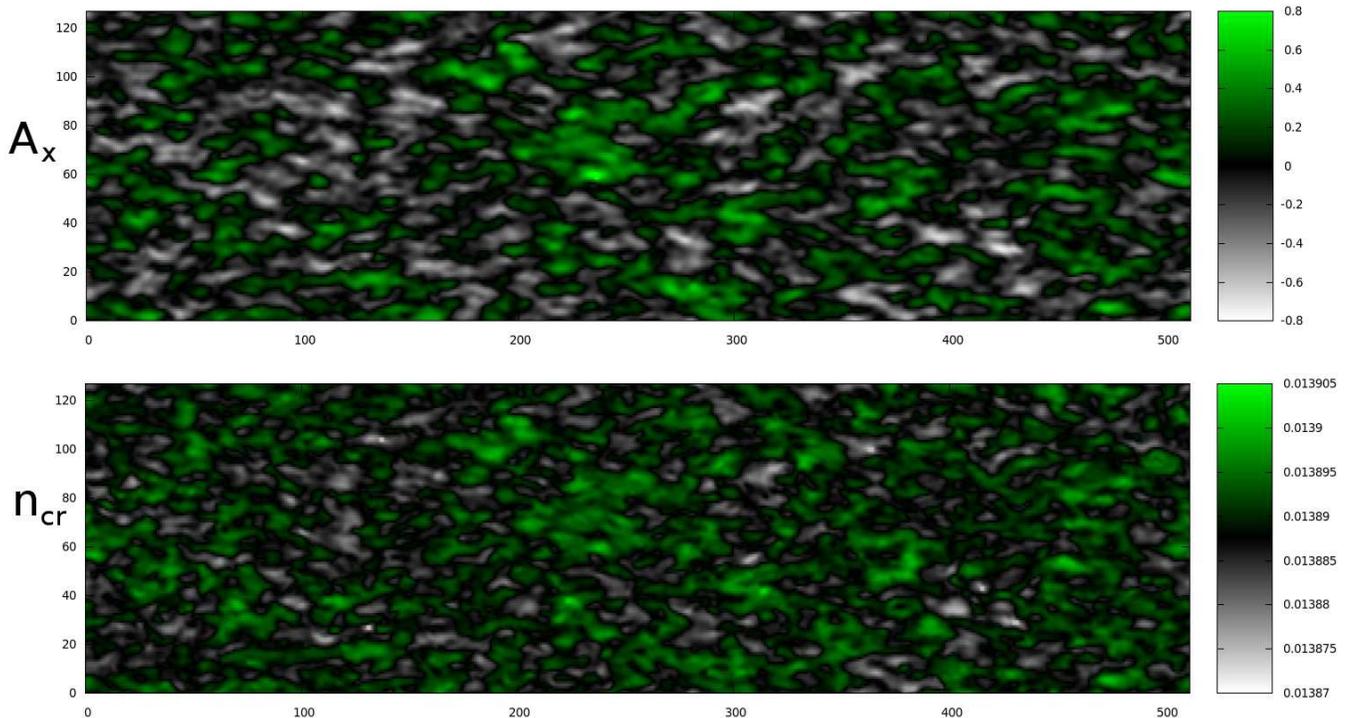}
\caption{Slice through the $x-y$ plane demonstrating the correlation 
between $A_x$ and $n_{\rm cr}$. While this is found to occur at early times, 
no additional focussing was found in any of the simulations, and the 
correlation eventually disappears in the non-linear stage.}
\label{fig:ANcomp} 
\end{figure*}

Considering very young supernovae, the results presented here 
suggest that shocks which are efficiently accelerating cosmic rays
might have an observable time dependent behaviour, as both the 
cosmic-ray efficiency and self-generated magnetic field fluctuations increase. 
SN1987a represents an interesting example, since the radio flux has been increasing 
steadily for over 20 years, implying an increasing acceleration efficiency, while the spectrum has 
been steadily flattening \citep{zanardoetal}. Reconciling this behaviour within the 
non-linear diffusive shock acceleration framework requires a somewhat 
arbitrary fine-tuning of the electron-proton injection ratio \citep{berezhko11}. 
However, this 
time-dependent behaviour is quite consistent with acceleration at an oblique shock.
Since the shock is believed to be expanding into a Parker spiral \citep{kirkwassmann},
at least for some of its lifetime,
the majority of the shock surface should have large magnetic obliquities. As 
the acceleration becomes more efficient, the magnetic fluctuations become amplified,
and the collision frequency increases. Comparing to figure 1 of \cite{Belletal11},
this will lead to a gradual flattening of the spectrum. As the field enters the 
non-linear regime of magnetic field amplification, the perpendicular currents
are damped near the shock, and a power-law spectrum closer to the 
standard test-particle result should result. While this explanation is clearly not in 
any way rigorous, it does provide an alternative to the non-linear
shock acceleration model, which neglects the effects of the obliquity of the upstream 
magnetic fields, and additionally any deviations from the diffusion approximation.

\subsection{Parallel shocks}

Parallel shocks have received the greatest attention in the literature, 
both in terms of particle acceleration and resulting magnetic field amplification.
We have performed an extensive suite of simulations investigating the growth of
magnetic field in this configuration, to investigate a number of effects.

Recently, \cite{revillebell12} presented a theoretical and numerical investigation
of cosmic-ray filamentation and its role in the generation of large scale 
magnetic fluctuations. The
results were based on a two-dimensional analysis that assumed slab-symmetry along
the direction of cosmic-ray anisotropy. It was demonstrated that the cosmic-ray 
density correlated with the component of the magnetic vector potential parallel to 
the direction of cosmic ray streaming. 

Using the VFP code, we are able to drop the approximation of slab symmetry
and investigate if the process occurs when the third dimension is
included. The code is set up as before, with periodic boundary conditions and 
a large scale imposed gradient to drive a current consistent with equation 
(\ref{CRfluxes}). It is found in all cases that the cosmic-ray density correlates 
with the vector potential during the linear growth ($\delta B \ll B_0$)
as can be seen in Figure \ref{fig:ANcomp}. 
In fact, the correlation also exists at oblique shocks, although the correlation
is between the cosmic-ray density and the component of the vector potential 
parallel to the mean cosmic-ray current. In this sense, the linear analysis 
of \cite{revillebell12} still applies, and one might expect growth 
of large scale fields. However, the resulting filaments that 
form do not continue to focus, and the structures are ultimately destroyed in 
the non-linear stage. It should be pointed out that we only consider the 
evolution of cosmic rays at a single energy, and that the simulation domain is 
periodic. It is still possible, and in fact quite likely, 
that the leading edge of the precursor can still filament
(see recent results from \cite{caprioli}). The periodic simulations 
presented here are more representative of a region deep in the precursor.
An extensive 
parameter scan failed to produce noticeable elongated structures 
in our simulations. On the other hand, the 
results do seem to suggest that cosmic rays can only escape the source at 
the highest energies, which is consistent with current models of cosmic-ray escape. 
Full, energy-dependent shock simulations are still required to rule out the 
possible significance of cosmic-ray filamentation as a low energy 
cosmic-ray escape channel.

\begin{figure}
 \includegraphics[width=0.48\textwidth]{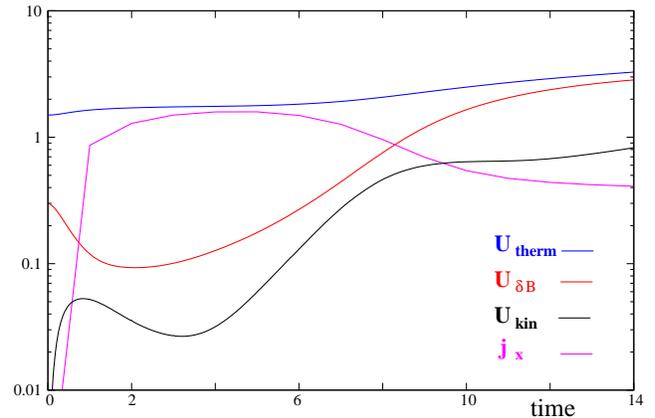}
\caption{Time evolution of the different components of the energy 
density. $U_{\delta B} = \langle B^2 - B_0^2 \rangle/8\pi$ is the 
energy density in the magnetic fluctuations having subtracted 
off the energy associated with the mean field. All energy densities 
are in units $B_0^2/8\pi$. (The sharp jump in 
$j_x$ from t=0 to t=1 is due to the fact that the current is only 
stored at integer timesteps.)}
\label{fig:growth}
\end{figure}

The simulations allow us also to address, the interaction between the cosmic rays and the 
self-consistently amplified
magnetic fields, over a spatial range that covers both the gyro-resonant interactions
and the sub-Larmor scale fluctuations where the fast growing NRH instability 
operates. A number of interesting effects are found.
First, it is noted that the evolution of the plasma is quite different 
from that typically seen in the fixed cosmic-ray current simulations 
\citep{bell04,revilleetal08,zirakashvilietal}. 
Figure \ref{fig:growth} shows the evolution of the 
different components of the energy. Previous MHD simulations of the field amplification 
using a constant uniform current typically go through three stages. 
An initial inertial phase, where the plasma is set in motion, and the magnetic 
field is either constant or can actually decrease as energy is transferred to the 
background fluid motions. Once in motion, the plasma starts to stretch the 
field lines, and reinforced by the $\bm{j}\times\bm{B}$ force, leads to a 
run-away instability. Finally, in the non-linear phase, the plasma motion is 
highly disordered and rotational. The kinetic energy continues to grow as
$t^2$ \citep{pelletieretal,zirakashvilietal}, and the magnetic energy 
grows in equipartition with the same scaling.

When we include the feedback on cosmic rays, the initial behaviour is the same, 
but the growth of kinetic energy stagnates at early times. The magnetic field 
starts to grow, with total energy density in the growing magnetic modes
quickly dominating the kinetic energy.  
Already at $\delta B^2/B_0^2 \sim 0.1$ the scattering on magnetic structures
acts to reduce the current such that the driving term can no longer compensate. 
Eventually, the cosmic-ray current becomes too weak to have any influence on 
the magnetic field on a timescale comparable with the precursor crossing time.

\begin{figure*}
 \includegraphics[width=\textwidth]{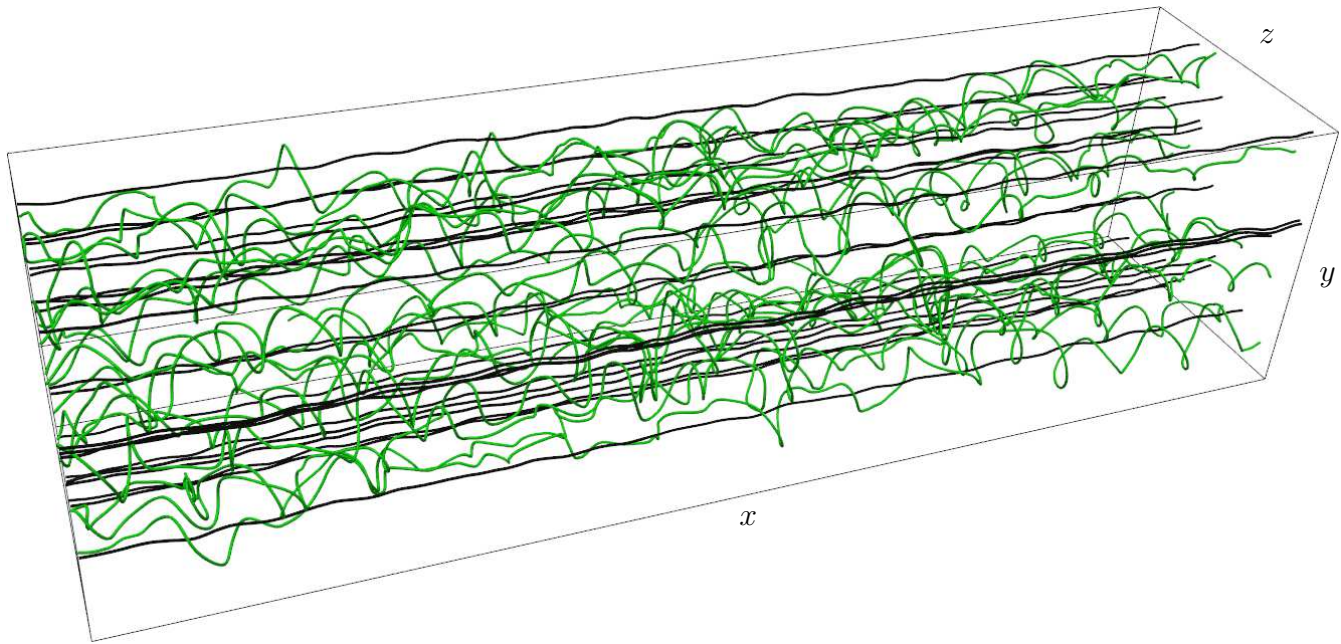}
\rput(1,2){{\Large $x$}}
\rput(8.7,5.2){{\Large $y$}}
\rput(7.9,8.4){{\Large $z$}}
\caption{Plot of the field lines for $\bm{j}$ and $\bm{B}$
from time $t=10$ for the simulation shown in Figure \ref{fig:growth}.
The approximately straight dark lines correspond to the current, while 
the green helical lines show the magnetic fieldlines. 
The gyroradius of the cosmic rays in the mean field
is $3/4$ the length of the simulation box.}
\label{fig:lines}
\end{figure*}

An interesting result of amplifying the magnetic field on sub-Larmor scales, 
is that the frequent scattering on the resulting non-linear fluctuations
prevents the current from following the field on large scales. This can clearly 
be seen in Figure \ref{fig:lines}, where the dominant magnetic field structures 
are on scales much less than the particle gyroradius which is $3/4$ the box 
length and 3 times the box width. Thus, one can safely assume that the usual 
expression for the linear growth rate of parallel modes ($\bm{k}\|\bm{B}_0$)
\eqb
\gamma = \sqrt{\frac{kj_{\rm cr} B_0}{\rho c}}
\eqe 
holds up to wavelengths larger than the gyroradius of the particles 
driving the growth. Of course, the simulations are initialised with a perfectly 
uniform mean field, and variations in the orientation of the 
large scale magnetic field in the pre-existing ambient circumstellar medium
may introduce interesting additional effects \citep{giacinti}. Nevertheless,
a faster growth rate at long wavelengths is clearly implied. 

The role of collisional effects has also been discussed 
in \cite{shurebell11}. Whether this is the same instability is uncertain,
since non-resonant scattering on short wavelength fluctuations appears to dominate, and 
unlike the modes analysed in \cite{shurebell11}, the growing waves are exclusively 
of a single polarisation, rotating in the opposite sense to the cosmic rays in the 
mean field, i.e. non-resonant.

\subsection{Regulated driving}
\label{sect:reg}

For the simulations reported in the previous sections, a fixed driving 
term, determined by the initial conditions and our numerical 
collision term was used (equation \ref{CRgrad}). Since the purpose of these
investigations is to study the interaction of particles with their own 
self-generated non-linear structures, we naturally expect a reduction in
the mean free path, and consequently, for the system we consider, 
a similar reduction in the current (Figure \ref{fig:growth}).

\begin{figure}
 \includegraphics[width=0.48\textwidth]{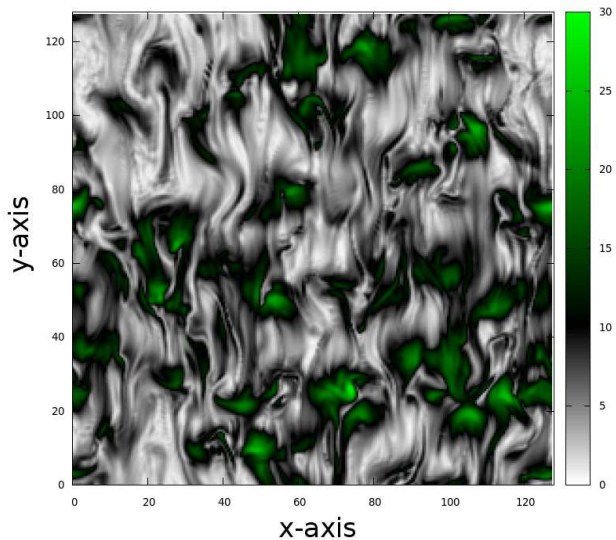}
\caption{Slice through the $x-y$ plane, showing the magnitude of the magnetic 
field, in units of $B_0$. The regions of intense magnetic field have an average 
strength of $\sim 25 B_0$ and a size $L\sim 10$. The gyroradius of the cosmic rays 
in the initial field is $r_{\rm g0} = 256$, such that 
these regions of intense field can deflect particles through 
$\Delta \theta \sim L/r_{\rm g}\sim 90^\circ$.}
\label{fig:blobs}
\end{figure}

\begin{figure}
 \includegraphics[width=0.48\textwidth]{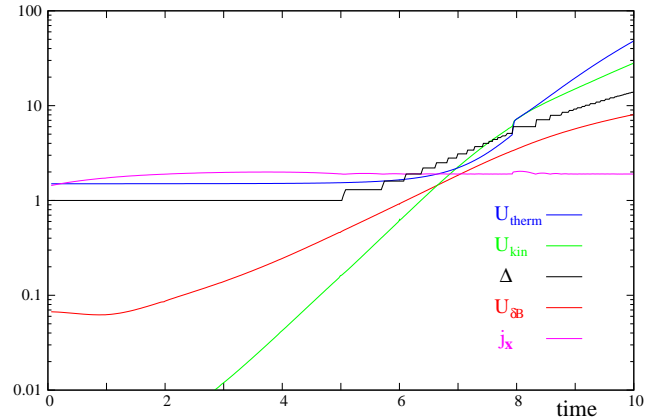}
\caption{Growth of the different components of the energy density 
as a function of time for simulations with regulated driving. The current 
density $j_x$ and $\Delta$ the renormalisation parameter of the driving
are also shown. $\nu = 0.125 \Omega_0$, $u_{\rm sh} = 0.1 c$.
Time units are $5.2/\Omega_0$} 
\label{fig:driven}
\end{figure}

At an actual (quasi-parallel) shock, the increased scattering rate would
prevent particles escaping far upstream of the shock, and the scaleheight of 
the precursor would decrease. Thus, it is reasonable to expect that the  
cosmic-ray gradient and the diffusion might conspire to maintain
the total current density at a constant level, i.e.
$$ j_{cr} = e\int \diff^3 p D \pdiff{f_0}{x} \approx {\rm constant}\enspace.$$
We have performed 
simulations where the large scale gradient (the driving term) is increased 
in such a way 
as to maintain an approximately constant cosmic-ray current as the field 
continues to grow. Starting from the usual result, we allow the current 
to build up to $t=3$, and store this value as a reference current. As the
magnetic fluctuations grow and dominate, the driving term is increased using 
a normalisation parameter $\Delta$. This parameter then becomes a measure of 
the effective reduction in the diffusion coefficient 
$\Delta = D_B/D$, where $D_B$ is the Bohm value determined from the initial 
conditions, which for the simulations shown has $\nu = \Omega/8$.  
It is found that, not only does the diffusion coefficient reduce well below the 
value imposed by the initial conditions (as expected), but by the end of the 
simulation, has even reduced below the Bohm limit in the pre-existing field 
$\nu_{\rm eff} \sim 2 \Omega$.
This can be clearly seen in Figure \ref{fig:driven}. The simulation is terminated 
when the maximum fast-magnetosonic velocity becomes comparable with the 
cell crossing time of the cosmic rays. This happens quite early,
due to the artificially large Alfv\'en velocity used in the simulations. 
In reality, the fields would continue to grow for considerably longer, as found
for example in \cite{bell04}, and the diffusion coefficient would likely continue 
to decrease in parallel. We note however, that the scattering is likely dominated 
by localised regions of very intense magnetic field. At the end of the simulation
the magnetic structures are sufficiently strong and on sufficient scale that 
they can deflect a cosmic ray through $\sim90^\circ$, 
as shown in Figure \ref{fig:blobs}. 

Since the maximum value of the magnetic field is 
in general time-limited (saturation of the magnetic field has not been found 
in any simulation with constant/sustained current), it is expected that the 
diffusion coefficient of the cosmic-rays will continue to decrease with growing 
magnetic field, presumably well beyond the level shown here. It is commonly
assumed that Bohm diffusion in the amplified magnetic field is an acceptable 
value for the diffusion coefficient in most calculations of relevance to
diffusive shock acceleration. Thus, it is important to be clear about
what this means. In section \ref{sims_sect}, the Bohm scaling argument was 
derived. Typically when referring to the Bohm limit, it is thought that the 
collision frequency matches the gyrofrequency, such that 
\eqb
D_{\rm B}^* = \frac{1}{3}\frac{v^2}{\Omega} = \frac{1}{3}r_{\rm g} v\enspace.
\eqe 
For highly disordered magnetic fields, confusion can now arise, as it is unclear
what value of the magnetic field should be taken when determining the 
gyroradius, and more importantly, if the Bohm limit itself applies in nature
in the presence of strong amplification.
In the simulations presented here, our experimentally determined diffusion 
coefficient is half that of the Bohm limit value if we adopt 
the gyroradius in the pre-amplified magnetic field, but roughly a factor of
four times the Bohm limit in the root mean squared field.\footnote{
This applies only to the particles driving the field growth within our 
mono-energetic approximation. Previous work by \cite{revilleetal08} have 
already demonstrated the energy dependence of the transport properties in 
self generated turbulence when the particles were all at considerably lower 
energies than those assumed driving the field growth.
} However, the diffusion 
coefficient was still reducing at the time the simulation was terminated, suggesting
that we might expect the diffusion coefficient at a supernova to correspond to 
its Bohm value in the amplified magnetic field, modulo a numerical factor of order
unity. Thus, it may not be such an extreme approximation when considering the 
maximum energies at young supernovae, to adopt the Bohm limit 
in the amplified magnetic field, namely the one inferred from observations.
Unfortunately, it remains uncertain as to whether the majority of the amplification 
occurs in the upstream, or acts simply as a seed for Richtmeyer-Meshkov 
type instabilities at the shock front itself \citep{giacjok, schurebell13}. However,
as noted by \cite{guoetal}, the growth in the downstream region due to such clumpy 
interaction occurs over a large distance,
and some additional mechanism is required to explain the strong fields observed 
close to the shock. This favours a scenario in which at least a sizeable fraction 
of the amplification occurs upstream. Future simulations that include a shock in the 
simulations will shed further light on this issue. 
  
Finally we note that young supernovae having clear Balmer emission from the 
outer shock may be used as an indirect probe of upstream amplification.
All of the simulations, independent of magnetic obliquity show a common 
feature, namely that in the non-linear regime,
the thermal energy density is greater than the magnetic energy
density by typically an order of magnitude, e.g. Figure \ref{fig:driven}. 
Thus, if fields of $100\mu$G are 
produced upstream, a thermal energy density in excess of 1~keV/cc may also be produced.
The survival of neutrals in such high temperature plasmas is highly unlikely, 
unless the precursor is very short. However, a short precursor will not provide  
sufficient time to amplify fields to such high values. Several young remnants 
show regions of both non-thermal x-ray filaments, synonymous with efficient particle 
acceleration, and broad/narrow H$\alpha$ emission, suggesting survival of neutrals
into the downstream \cite[e.g.][]{winkler,ghavamian,helder}. If indeed 
the magnetic fields are produced predominantly upstream, these regions should 
anti-correlate, thus providing indirect evidence of in-situ acceleration of cosmic
ray protons or nuclei, or indeed the lack thereof. Further observations are 
necessary to answer this question.

\section{Discussion}
\label{disc_sect}

We have presented a series of numerical simulations investigating the 
interaction between cosmic rays and magnetic fields in the precursors of
supernova remnants. The investigations are carried out using a hybrid 
MHD-kinetic code which applies a spherical harmonic expansion of the 
Vlasov-Fokker-Planck equation. This code allows us to consider a range 
of multi-scale phenomena previously inaccessible to numerical simulations. 
A number of results have followed from these simulations that can be applied 
to existing theories of particle acceleration, and in particular the 
observational signatures of efficient acceleration at young supernova remnants.

Previous work by \cite{revillebell12} proposed a method whereby low energy 
cosmic rays could escape the remnant via self-focussing into low density 
cavities surrounded by strong magnetic fields. The analysis and simulations 
performed in \cite{revillebell12} were exclusively in a 2D slab geometry, which
limited the amount of focusing that could be achieved. This is a consequence 
of the conservation of the component of the vector potential into the plane
in such a geometry.  
Including the third dimension allows for growth of this component, and in 
theory, continued focusing, which could result in formation of filamentary 
structures aligned along the direction 
of cosmic-ray streaming. The current 3D simulations appear to support 
the linear analysis of \cite{revillebell12}, but that the resulting structures 
become disrupted over a short distance. While this disfavours the 
possibility of non-diffusive escape channels for cosmic rays from the 
remnants, it does provide convincing evidence that the escape of cosmic rays 
into the interstellar medium at quasi-parallel shocks is limited to a narrow 
window around the maximum energy, consistent with current models.
    
Oblique and perpendicular shocks present a slightly different view, 
although there is clear evidence for a universal behaviour closer to the shock.  
Far upstream of the shock, where the scattering is weak, particles are tied to
field-lines, and advected back towards the shock. It is demonstrated however, that
the self-generated fields closer to the shock act to damp any 
diamagnetic-currents and in a sense, behave very much like parallel shocks.
At the same time, considerable amplification of the magnetic field can occur.
Previous suggestions that quasi-perpendicular shocks are faster accelerators 
have implicitly assumed that the scattering was weak \citep{jokipii87}.
Later calculations demonstrated that the maximum energy became 
a problem of geometry rather than time \citep{achterberghouche,bell08}, as the 
energisation 
relies on drifting along the shock surface. On the global scale of
the precursor, we now have a rather unusual situation, 
quite distinct from parallel shocks, 
where the particles can diffuse isotropically close to the shock,
but are strongly confined by the upstream undisturbed field. Whether this acts 
to increase or decrease the maximum energy is not immediately clear. However,
full shock simulations will be performed to answer this question in the near future.

Oblique shocks may also play an important role in explaining observations 
from young supernovae, or likewise, early observations can provide important
information on the acceleration process. Immediately after break-out, 
supernovae typically expand 
into a progenitor wind. It is usually assumed in such winds that the radial 
magnetic field falls off quite rapidly, while the perpendicular component
will persist. Thus young supernovae provide ideal laboratories to investigate 
many of the ideas presented here. We have demonstrated, albeit in a very qualitative 
sense, how the current evolution of SN 1987a can be explained within the 
framework of acceleration at highly oblique shocks without the need for 
non-linear cosmic-ray pressure feedback.  

The possibility of sub-Bohm diffusion (with respect to the pre-existing ambient field
strength) of cosmic rays at the highest energies has been demonstrated 
self-consistently for parallel shocks, under the assumption that the driving 
regulates itself. On the other hand, the normal component of the diffusion 
tensor at oblique shocks is initially sub-Bohm, but tends to increase due to 
amplification of magnetic perturbations. 
Regulation is more complicated in these geometries, and full shock simulations  
are required. However, the high resolution simulations presented in this paper,
show that the self-generated fields naturally produce highly disordered field lines
strongly suggesting a universal behaviour of isotropic diffusion at shocks that 
are efficiently accelerating cosmic rays.

\section*{Acknowledgements}

{ The research leading to these results has received funding from the European Research Council under the 
European Community's Seventh Framework Programme (FP7/2007-2013) / ERC grant agreement no. 247039. 
BR thanks M. Tzoufras and C. Ridgers for valuable discussions on VFP codes, and G. Giacinti for 
helpful comments on the paper.}

\bibliographystyle{mn2e}

\appendix
\onecolumn

\section[]{Spherical harmonic expansion of Vlasov-Fokker-Planck equation}

In the following, we choose the polar axis to lie along the $x$-direction, 
such that the Cartesian momentum vector expressed in terms of
spherical angles is
$\bm{p}=\left(p\cos\theta, p\sin\theta\cos\phi,p\sin\theta\sin\phi\right)$.
Using the recursion relations and orthogonality condition of the spherical 
harmonics, the time derivative
for each component of the expansion can be written in the form
\eqb
\label{eqn:A1}
\pdiff{f_\ell^m}{t}-U_{x,\ell}^m - U_{y,\ell}^m- U_{z,\ell}^m -
A_{x,\ell}^m - A_{y,\ell}^m- A_{z,\ell}^m-B_{x,\ell}^m - B_{y,\ell}^m
- B_{z,\ell}^m-E_{x,\ell}^m - E_{y,\ell}^m- E_{z,\ell}^m-
\sum_{x,y,z}T^m_{\alpha\beta,\ell}=C_\ell^m
\eqe

Most of these terms have been previously given in \cite{kalos} and
\cite{tzoufras}, but are repeated here so all the equations can be 
found in one place. The mixed 
coordinate frame introduces several additional terms, contained in the 
$T^m_{\alpha\beta,\ell}$, which can be associated with the adiabatic term 
in equation (\ref{vfp_local}) in the main text. The subscripts 
$\alpha,\beta\in \lbrace x,y,z\rbrace$ correspond to the
relevant term containing the partial derivative 
$\partial u_\alpha/\partial x_\beta$. Although cumbersome,
the derivation of each term is straightforward, and the final expression often displays 
a noticeable symmetry, a clear indication of the applicability of  
spherical harmonics in solving problems of this nature. 

The advection terms are as follows
\eqb
\label{eq:advect}
U_{i,\ell}^m&=&-u_i\pdiff{f_\ell^m}{x_i} \enspace 
\forall \enspace i\in\lbrace x,y,z\rbrace
\nonumber\\
A_{x,\ell}^m&=&-v\pdiff{}{x}\left[\frac{\ell-m}{2\ell-1}f_{\ell-1}^m+\frac{\ell+m+1}{2\ell+3}f_{\ell+1}^m\right]
\nonumber\\
A_{y,\ell}^{m>0}+A_{z,\ell}^{m>0}&=& -\frac{v}{2}{\rm D}^+\left[ 
\frac{(\ell-m)(\ell-m-1)}{2\ell-1}f_{\ell-1}^{m+1}-
\frac{(\ell+m+1)(\ell+m+2)}{2\ell+3}f_{\ell+1}^{m+1}\right]\nonumber\\
&&+\frac{v}{2}{\rm D}^-\left[ 
\frac{1}{2\ell-1}f_{\ell-1}^{m-1}-
\frac{1}{2\ell+3}f_{\ell+1}^{m-1}\right]\nonumber\\
A_{y,\ell}^{0}+A_{z,\ell}^{0}&=&-\Re\left\lbrace v{\rm D}^+\left[ 
\frac{\ell(\ell-1)}{2\ell-1}f_{\ell-1}^{1}-
\frac{(\ell+1)(\ell+2)}{2\ell+3}f_{\ell+1}^{1}
\right]\right\rbrace
\eqe
where $D^\pm=\frac{\partial \,}{\partial y}\pm\im\frac{\partial \,}{\partial z}$.

Defining the cyclotron rotation vector $\bm{\Omega}=q\bm{B}/\gamma m c$, 
with $q$ the particle's charge and $\gamma m$ its (relativistic) mass, the 
rotational terms associated with the magnetic field components are
\eqb
B_\ell^{m>0}&=&\im\Omega_x m f_\ell^m+\frac{1}{2}\left[ 
(\Omega_z+\im\Omega_y)f_{\ell}^{m-1}-
(\ell+m+1)(\ell-m)(\Omega_z-\im\Omega_y)f_{\ell}^{m+1}
\right]\nonumber\\
B_\ell^{0}&=&-\ell(\ell+1)\Re\left[ 
(\Omega_z-\im\Omega_y)f_{\ell}^{1}
\right]
\eqe
The spherical harmonic expansion has the additional feature that the rotational terms
are algebraic, and do not involve partial derivatives, as would be the case 
for a Cartesian tensor expansion \citep{johnston}.

Recalling that in the mixed coordinate frame, the local acceleration has an
equivalent form to the electric field in the standard Vlasov equation, we 
thus use the notation
\eqbn
\bm{E}=\frac{1}{c}\left[\pdiff{\bm{u}}{t}+(\bm{u}\cdot\bm{\nabla})\bm{u}\right],
\eqen
from which it follows
\eqb
E_{x,\ell}^m&=&E_xp\left[\frac{\ell-m}{2\ell-1}G_{\ell-1}^m
+\frac{\ell+m+1}{2\ell+3}H_{\ell+1}^m\right]\nonumber\\ 
{E}_{y,\ell}^{m>0}+{E}_{z,\ell}^{m>0} 
&=&-\frac{p}{2}\biggl[\frac{E_y-iE_z}{2\ell-1}G_{\ell-1}^{m-1}-
 \frac{E_y+iE_z}{2\ell-1}(\ell-m)(\ell-m-1)G_{\ell-1}^{m+1}\nonumber\\
& & 
-\frac{E_y-iE_z}{2\ell+3}H_{\ell+1}^{m-1}+ \frac{E_y+iE_z}{2\ell+3}(\ell+m+1)(\ell+m+2)
H_{\ell+1}^{m+1}\biggr] \nonumber
\\
{E}_{y,\ell}^{0}+{E}_{z,\ell}^{0} 
&=&\Re\biggl\{p(E_y+iE_z)\biggl[\frac{\ell(\ell-1)}{2\ell-1}G_{\ell-1}^1-
\frac{(\ell+1)(\ell+2)}{2\ell+3}H_{\ell+1}^1\biggr]\biggr\}
\eqe
where
\eqb
\label{eq:GandH}
G_\ell^m=p^\ell\pdiff{}{p}\left(p^{-\ell}f_\ell^m\right), \enspace
H_\ell^m=p^{-(\ell+1)}\pdiff{}{p}\left(p^{(\ell+1)}f_\ell^m\right)
\eqe

Since the adiabatic term has two occurences of the momentum vector, it contains 
nine different terms associated with the different combinations of the axes.
However, due to the symmetry about the polar axis, most of these terms are
very similar and can be grouped together. Some of these terms have been previously 
derived in one \citep{Belletal11} and two dimensions 
(C. Ridgers private communication).
\eqb
T_{xx,\ell}^m &=& p\pdiff{u_x}{x}\biggl[
\frac{(\ell-m)(\ell-m-1)}{(2\ell-3)(2\ell-1)}G_{\ell-2}^m
+\frac{(\ell-m+1)(\ell+m+1)}{(2\ell+3)(2\ell+1)}G_{\ell}^m\nonumber\\
&&+\frac{(\ell+m)(\ell-m)}{(2\ell+1)(2\ell-1)}H_{\ell}^m 
+\frac{(\ell+m+1)(\ell+m+2)}{(2\ell+3)(2\ell+5)}H_{\ell+2}^m\biggr]\nonumber
\\
T_{xy,\ell}^{m>0}+T_{xz,\ell}^{m>0}&=&\frac{p}{2}\left(\pdiff{u_x}{y}-\im\pdiff{u_x}{z}\right)
\biggl[\frac{\ell+m+1}{(2\ell+5)(2\ell+3)}H_{\ell+2}^{m-1}\nonumber\\
&&+\frac{\ell-m+2}{(2\ell+1)(2\ell+3)}G_{\ell}^{m-1}-
\frac{\ell+m-1}{(2\ell+1)(2\ell-1)}H_{\ell}^{m-1}
-\frac{\ell-m}{(2\ell-1)(2\ell-3)}G_{\ell-2}^{m-1}\biggr]\nonumber
\\
&&+\frac{p}{2}\left(\pdiff{u_x}{y}+\im\pdiff{u_x}{z}\right)\biggl[
-\frac{(\ell+m+1)(\ell+m+2)(\ell+m+3)}{(2\ell+5)(2\ell+3)}H_{\ell+2}^{m+1}\nonumber\\
&&-\frac{(\ell-m)(\ell+m+1)(\ell+m+2)}{(2\ell+1)(2\ell+3)}G_{\ell}^{m+1}+
\frac{(\ell-m)(\ell+m+1)(\ell-m-1)}{(2\ell+1)(2\ell-1)}H_{\ell}^{m+1}\nonumber \\
&&+\frac{(\ell-m)(\ell-m-1)(\ell-m-2)}{(2\ell-1)(2\ell-3)}G_{\ell-2}^{m+1}\biggr]\nonumber
\\
T_{xy,\ell}^{0}+T_{xz,\ell}^{0}&=&p\Re\biggl\lbrace\left(\pdiff{u_x}{y}+\im\pdiff{u_x}{z}\right)
\biggl[
-\frac{(\ell+1)(\ell+2)(\ell+3)}{(2\ell+5)(2\ell+3)}H_{\ell+2}^{1}\nonumber\\
&&-\frac{\ell(\ell+1)(\ell+2)}{(2\ell+1)(2\ell+3)}G_{\ell}^{1}+
\frac{\ell(\ell+1)(\ell-1)}{(2\ell+1)(2\ell-1)}H_{\ell}^{1}
+\frac{\ell(\ell-1)(\ell-2)}{(2\ell-1)(2\ell-3)}G_{\ell-2}^{1}\biggr]\biggr\rbrace\nonumber
\eqe

\eqb
T_{yx,\ell}^{m>0}+T_{zx,\ell}^{m>0}&=&\frac{p}{2}\partial_x\left(u_y-\im u_z\right)
\biggl[\frac{\ell+m+1}{(2\ell+5)(2\ell+3)}H_{\ell+2}^{m-1}\nonumber\\
&&-\frac{\ell+m+1}{(2\ell+1)(2\ell+3)}G_{\ell}^{m-1}+
\frac{\ell-m}{(2\ell+1)(2\ell-1)}H_{\ell}^{m-1}
-\frac{\ell-m}{(2\ell-1)(2\ell-3)}G_{\ell-2}^{m-1}\biggr]\nonumber
\\
&&+\frac{p}{2}\partial_x\left(u_y+\im u_z\right)\biggl[
-\frac{(\ell+m+1)(\ell+m+2)(\ell+m+3)}{(2\ell+5)(2\ell+3)}H_{\ell+2}^{m+1}\nonumber\\
&&+\frac{(\ell-m)(\ell+m+1)(\ell-m+1)}{(2\ell+1)(2\ell+3)}G_{\ell}^{m+1}-
\frac{(\ell-m)(\ell+m+1)(\ell+m)}{(2\ell+1)(2\ell-1)}H_{\ell}^{m+1}\nonumber \\
&&+\frac{(\ell-m)(\ell-m-1)(\ell-m-2)}{(2\ell-1)(2\ell-3)}G_{\ell-2}^{m+1}\biggr]\nonumber
\\
T_{yx,\ell}^{0}+T_{zx,\ell}^{0}&=&p\Re\biggl\lbrace\partial_x\left(u_y+\im u_z\right)
\biggl[
-\frac{(\ell+1)(\ell+2)(\ell+3)}{(2\ell+5)(2\ell+3)}H_{\ell+2}^{1}\nonumber\\
&&+\frac{\ell(\ell+1)^2}{(2\ell+1)(2\ell+3)}G_{\ell}^{1}-
\frac{\ell^2(\ell+1)}{(2\ell+1)(2\ell-1)}H_{\ell}^{1}
+\frac{\ell(\ell-1)(\ell-2)}{(2\ell-1)(2\ell-3)}G_{\ell-2}^{1}\biggr]\biggr\rbrace\nonumber
\eqe

\eqb
T_{yy,\ell}^{m}+T_{yz,\ell}^{m}+T_{zy,\ell}^{m}+T_{zz,\ell}^{m}&=&
Q_{\ell}^{m}+R_{\ell}^m+S_{\ell}^{m}\nonumber
\eqe
where
\eqb
Q_{\ell}^{m>1}&=&pY\biggl[\frac{1}{(2\ell+3)(2\ell+5)}H_{\ell+2}^{m-2}-\frac{1}{(2\ell-1)(2\ell+1)}H_{\ell}^{m-2}
-\frac{1}{(2\ell+1)(2\ell+3)}G_{\ell}^{m-2}
+\frac{1}{(2\ell-3)(2\ell-1)}G_{\ell-2}^{m-2}\biggr]\nonumber
\\
S_{\ell}^{m}&=&pY^*\biggl[\frac{(\ell+m+1)(\ell+m+2)(\ell+m+3)(\ell+m+4)}{(2\ell+3)(2\ell+5)}H_{\ell+2}^{m+2}
-\frac{(\ell-m)(\ell-m-1)(\ell+m+1)(\ell+m+2)}{(2\ell-1)(2\ell+1)}H_{\ell}^{m+2}\nonumber\\
&&-\frac{(\ell-m)(\ell-m-1)(\ell+m+1)(\ell+m+2)}{(2\ell+1)(2\ell+3)}G_{\ell}^{m+2}
+\frac{(\ell-m)(\ell-m-1)(\ell-m-2)(\ell-m-3)}{(2\ell-3)(2\ell-1)}G_{\ell-2}^{m+2}
\biggr]\nonumber\\
R_{\ell}^m&=&\frac{p}{2}\left(\pdiff{u_y}{y}+\pdiff{u_z}{z}\right)
\biggl[\frac{(\ell+1)(\ell+2)+m^2}{(2\ell+1)(2\ell+3)}G_{\ell}^{m}
+\frac{\ell(\ell-1)+m^2}{(2\ell+1)(2\ell-1)}H_{\ell}^{m}\nonumber\\
&&-\frac{(\ell-m)(\ell-m-1)}{(2\ell-1)(2\ell-3)}G_{\ell-2}^{m}
-\frac{(\ell+m+1)(\ell+m+2)}{(2\ell+5)(2\ell+3)}H_{\ell+2}^{m}\biggr]
+\im\frac{p}{2}\left(\pdiff{u_y}{z}-\pdiff{u_z}{y}\right)
\frac{m}{2\ell+1}\left[G_\ell^m-H_\ell^m\right]\nonumber
\eqe
where
$$
Y=\frac{1}{4}\left(\pdiff{u_y}{y}-\pdiff{u_z}{z}\right)-\im
\frac{1}{4}\left(\pdiff{u_y}{z}+\pdiff{u_z}{y}\right)
$$

For $m<2$ we have $Q_{\ell}^0=(S_{\ell}^0)^*$ and 
\eqbn
Q_{\ell}^{1}=pY\biggl[\frac{\ell(\ell+1)}{(2\ell-1)(2\ell+1)}(H_{\ell}^{1})^*
-\frac{(\ell+2)(\ell+3)}{(2\ell+3)(2\ell+5)}(H_{\ell+2}^{1})^*
+\frac{\ell(\ell+1)}{(2\ell+1)(2\ell+3)}(G_{\ell}^{1})^*
-\frac{(\ell-2)(\ell-1)}{(2\ell-3)(2\ell-1)}(G_{\ell-2}^{1})^*\biggr]
\eqen

Finally, since the spherical harmonics are themselves solutions to Laplace's equation,
the collision term takes a particularly simple form
\eqb
\label{app:C}
C_\ell^m&=&-\frac{\nu}{2}\ell(\ell+1)f_{\ell}^m
\eqe

\bsp

\label{lastpage}

\end{document}